\documentclass[nonatbib]{ragtime} 

\title[Flares from spiral waves by lensing and time-delay amplification?]%
      {Flares from spiral waves by lensing and time-delay amplification?}

\author[V.~Karas, M.~Dov\v{c}iak, A.~Eckart and L.~Meyer]  
       {Vladim\'{\i}r Karas\at{1} 
        Michal Dov\v{c}iak\at{1}    
        Andreas Eckart\at[]{2,3}\splitauthors
	and Leonhard Meyer\at[]{2}\\
        \ins{1}Astronomical~Institute, Academy~of~Sciences, Bo\v{c}n\'{\i}~II~1401,\splitins[1]
        CZ-14131~Prague, Czech~Republic\\
        \ins{2}I.Physikalisches Institut, Universit\"at zu K\"oln, Z\"ulpicher~Str.~77,\splitins[1] D-50937~K\"oln, Germany\\
	\ins{3}Max-Planck-Institut f\"ur Radioastronomie, Auf~dem~H\"ugel~69,\splitins[1] D-53121~Bonn, Germany\\
}

\begin{document}


\begin{abstract}
Episodically accreting black holes are thought to produce flares when a
chunk of particles is accelerated to high velocity near the black hole
horizon. This also seems to be the case of Sagittarius~A* in the
Galactic Center, where the broad-band radiation is produced, likely via
the synchrotron self-Compton mechanism. It has been proposed that
strong-field gravitational lensing magnifies the flares. The effect of
lensing is generally weak and requires a fine-tuned geometrical
arrangement, which occurs with only a low probability. However, there
are several aspects that make Sagittarius~A* a promising target to
reveal strong gravity effects. Unlike type II (obscured) active
galaxies, chances are that a flare is detected at high inclination,
which would be favourable for lensing. Time delays can then
significantly influence the observed flare duration and the form of
light-curve profiles. 

Here we discuss an idea that the impact of lensing amplification should
be considerably enhanced when the shape of the flaring clump is
appropriately elongated in the form of a spiral wave or a narrow
filament, rather than a simple (circular) spot which we employed
previously within the phenomenological `orbiting spot model'. By
parameterizing the emission region in terms of the spiral shape and
contrast, we are able to extend the spot model to more complicated
sources. In the case of spirals, we notice a possibility that more
photons reach a distant observer at the same moment because of 
interplay between lensing and light-travel time. The effect is not
symmetrical with respect to leading versus trailing spirals, so in
principle the source geometry can be constrained. In spite of this, the
spot model seems to provide entirely adequate framework to study the
currently available data.
\end{abstract}


\section{Introduction}\label{intro}
Temporal changes of the radiation flux are frequently reported in
active galactic nuclei (AGN) as well as Galactic black-hole candidates,
i.e., two categories of objects that contain accreting black holes (for
reviews of AGN variability, see e.g.\ McHardy \& Czerny 1987; Lawrence
\& Papadakis 1993; Done 2002; Gaskell \& Klimek 2003; Vaughan, Fabian \&
Iwasawa 2005). Especially X-rays vary a lot and on short time-scales.
Variability time-scales extend down to the shortest resolvable intervals
and seem to scale with the black hole mass (Mirabel \& Rodr\'{\i}guez
1998; Papadakis 2004; Done \& Gierli\'nski 2005). Persisting
fluctuations are occasionally dominated by more substantial increases
of the radiation flux. These events have been dubbed `flares' and often
attributed to instabilities/shocks operating in black-hole accretion
flows (Haardt, Maraschi \& Ghisellini 1994; Poutanen \& Fabian 1999;
\.Zycki 2002; Czerny et al.\ 2004; Goosmann et al.\ 2006). One can
expect that fast bulk (orbital) motion and lensing play a role in
amplification of the flaring signal.

In the dynamical center of our Galaxy, a peculiar radio source,
Sagittarius~A* (Sgr~A*), is located (e.g.\ Eckart, Sch\"odel \&
Straubmeier 2005). It is very compact and presumably contains a
supermassive black hole. Given a relatively small distance
($D\simeq8$~kpc) and a large mass of the black hole
($M_{\bullet}\simeq3$--$4\times10^6M_{\odot}$), a silhouette of Sgr~A* should draw
a circle of diameter $\simeq10.4r_{\rm{}g}/D\simeq42\mu$arcsec on
the sky. Furthermore, a gaseous torus is not detected in Sgr~A*, and so
the central region can be viewed at high inclination, something which is
quite impossible in obscured AGN.

In spite of a very low level of its activity, flares of duration
$\simeq{}t_{\rm{}K}(r_{\rm{}ms})$ have been reported also from the
Galactic Center (Baganoff et al.\ 2001; Genzel et al.\ 2003; Marrone et
al.\ 2006; B\'elanger et al.\ 2005, 2006). Duration of short flares is
comparable with Keplerian orbital period near the  marginally stable
orbit, $t_{\rm{}K}(r_{\rm{}ms})$, and it is not much longer than the
light-crossing time across one gravitational radius:
$t_{\rm{c}}\,\equiv\,r_{\rm{}g}/c$,
$r_{\rm{g}}\,\equiv\,GM_{\bullet}/c^2\approx1.5\times10^{11}\,M_6\,\rm{cm}$,
$M_6\,\equiv\,M_{\bullet}/10^6M_{\odot}$. 

The flares occur about once per day from within a few milli-arcseconds
of Sgr~A* radio position. Because of short time-scales they cannot be
explained in terms of viscous processes in the standard accretion disc
with some appreciable accretion rate (as already mentioned, there is no
evidence for a standard-type axially symmetric accretion regime); Sgr~A*
is accreting at a highly sub-Eddington rate. Nonetheless, recent
millimeter, infrared and X-ray observations have confirmed these
irregular outbursts lasting between $\simeq20$ minutes and about
$2$~hours. They are probably generated by relativistic acceleration of
electrons in the innermost region, where synchrotron radiation emerges
followed by inverse Compton mechanism (Markoff et al.\ 2001; Yuan et
al.\ 2003; Liu, Melia \& Petrosian 2006). There are indications for
$17$--$20$~min quasi-periodicities to be present in light curves of some
of these flares (Genzel et al.\ 2003; Eckart et al.\ 2004). The
influences of relativistic lensing and Doppler effects have been
considered in connection with Sgr~A* since more than a decade ago (e.g.\
Hollywood \& Melia 1995; Melia et al.\ 2001). These effects are now of 
imminent interest because of growing amount of new data gathered in 
different wavebands.

The model of a bright spot orbiting near a black hole (Cunningham \&
Bardeen 1972, 1973; Bao \& Stuchl\'{\i}k 1992; Karas et al.\ 1992) has
been fairly successful in explaining the observed Sgr~A* modulation
(Broderick \& Loeb 2005, 2006; Meyer et al.\ 2006a,b; Noble et al.\
2007). It has been argued that the flare lightcurves can be understood
as a region of enhanced emission, a.k.a. `spot', that performs a
co-rotational bulk motion near above the innermost stable orbit,
$r=r_{\rm{}ms}$. The observed signal is modulated by relativistic
effects. According to this idea, Doppler and gravitational lensing
influence the observed radiation flux and this can be computed by
ray-tracing methods (Dov\v{c}iak et al.\ 2004a,b).

The original idea and the interpretation of the ``spot'' origin
have to be adapted to the conditions appropriate for Sgr~A*. To this
aim, the phenomenological model of the source is a way of parametrizing
the intensification of the signal. This approach can be extended to more
complicated geometry of the emission region, like standing shocks and
spiral waves, which is what we discuss here. For example, spiral waves
as an agent of light modulation have discussed and compared with the
spot model by Varni\`ere \& Blackman (2005) in the context of
quasi-periodic oscillations from accretion discs. On a more physical
level it is still not possible to calculate the intrinsic emissivity
from first principles, i.e., without enlarging the number of free
parameters beyond and reasonable limit. 

\section{Time delays from Sgr~A* vicinity}
\subsection{Model setup}
It is quite likely that the geometrical shape of the flare emission
region is deformed by shearing due to strong tidal fields of the black
holes, magnetohydrodynamic instabilities operating in the plasma, as
well as by the influence of stars passing nearby. Under such
circumstances the emission area can be better described as a transient
pattern extending in both radial and azimuthal directions. Relativistic
effects from spiral waves and standing shocks have been previously
invoked to explain spectral features from black-hole accretions discs
(Karas, Martocchia \& \v{S}ubr 2001; Hartnoll \& Blackman 2002; Machida
\& Matsumoto 2003; Fukumura \& Tsuruta 2004). Although the
Doppler boosting is visible even at a moderate value of the inclination
angle, much stronger enhancement can occur via gravitational lensing,
provided that a rather precise geometrical alignment with the caustic
position is satisfied (e.g.\ Rauch \& Blandford 1994; Bozza et al.\
2005).

It has been proposed that a kind of this instability could play a role
in forming Sgr~A* flares (Tagger et al.\ 1990, 2006) and since then the
idea of spiral perturbations has been greatly advanced (Falanga et al.\
2007).  Here we put forward a simple argument (based on Karas et al.\
2001) that relativistic effects together with finite light travel time
from different elements of the spiral source may add up together and
enhance the observed flare signal from Sgr~A*. For suitable spiral
shapes, $r\,\equiv\,r(\phi)$, the enhancement can reach quite
significant levels.
Lightcurve profiles depend on observer inclination, $\theta_{\rm{}o}$,
and the emission radius as the principal parameters, which in turn may
depend on the black hole spin $a$ through $r_{\rm{}ms}(a)$ dependency.

\begin{figure}
\begin{center}
\includegraphics[width=0.7\textwidth]{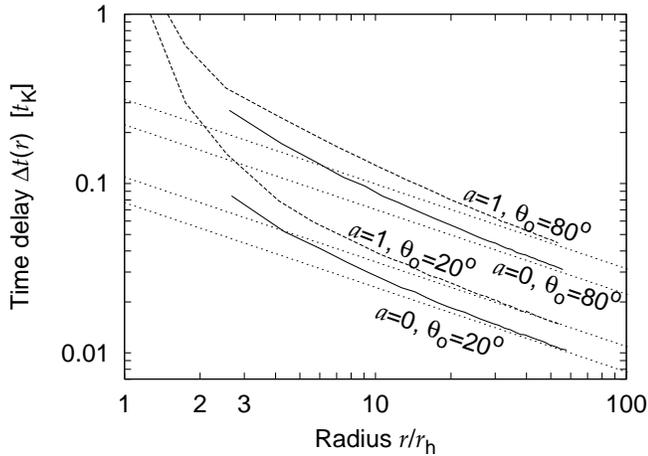}
\end{center}
\caption{Graphs of the maximum time delay ${\Delta}t(r)$ for photons
originating from an $r={\rm{}const}$ ring rotating in the black-hole
equatorial plane. The delay is plotted in units of the Keplerian
orbital period, $t_{\rm{}k}\simeq31(r^\frac{3}{2}+a)M_6$~sec. Radius is
expressed in $r_{\rm{}h}=[1+\sqrt{1-a^2}]^{1/2}r_{\rm{}g}$. Four cases
are shown with different spin $a$ of the black hole and inclination
$\theta_{\rm{}o}$ of the observer. The Euclidean estimate is plotted by
dotted lines of $-1/2$ slope. Towards low radius the relativistic delay
grows more rapidly than the estimate because of fast motion and strong
gravity.
\label{fig1}}
\end{figure}

\begin{figure}
\begin{center}
\includegraphics[width=0.48\textwidth]{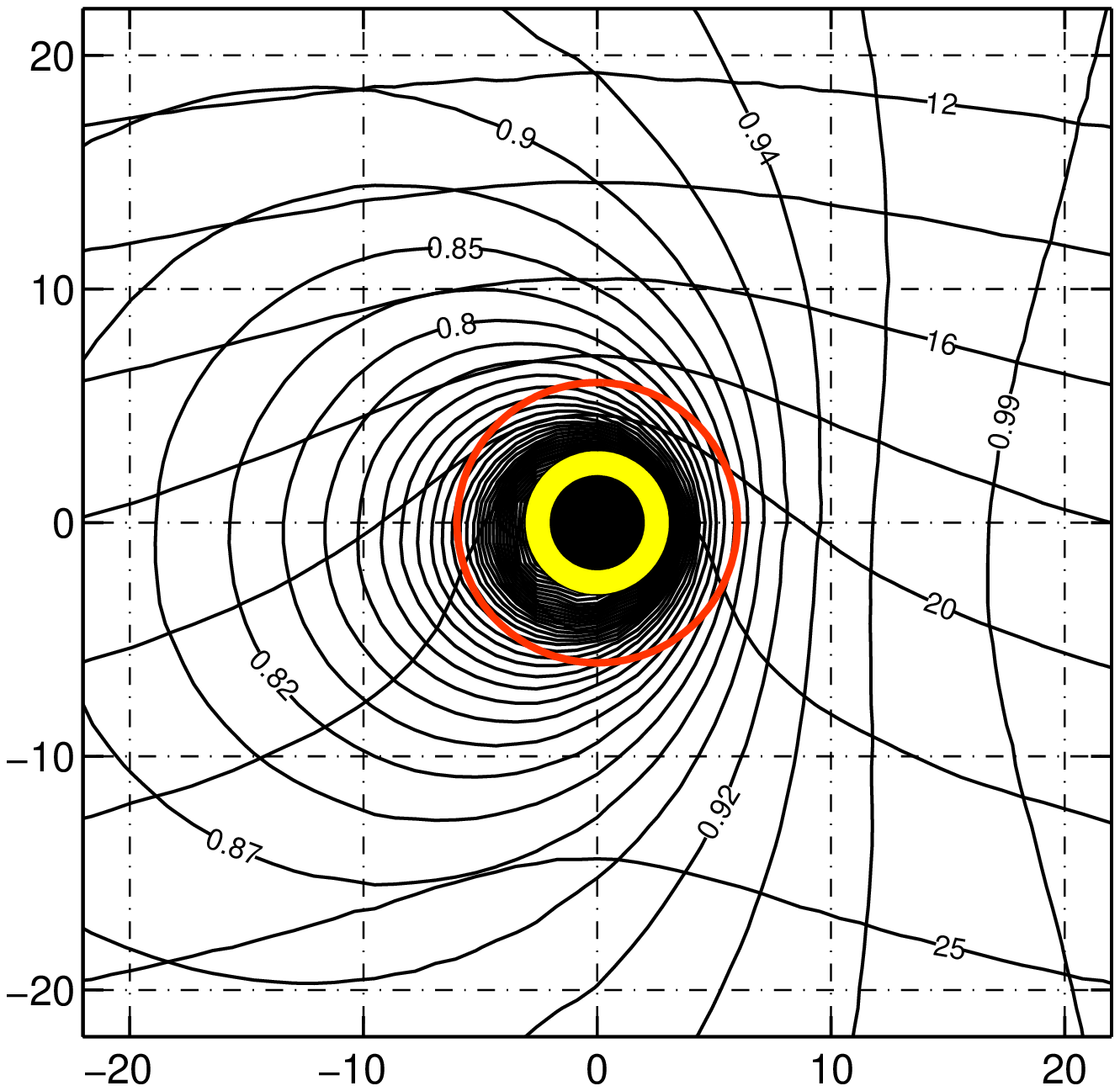}
\hfill
\includegraphics[width=0.48\textwidth]{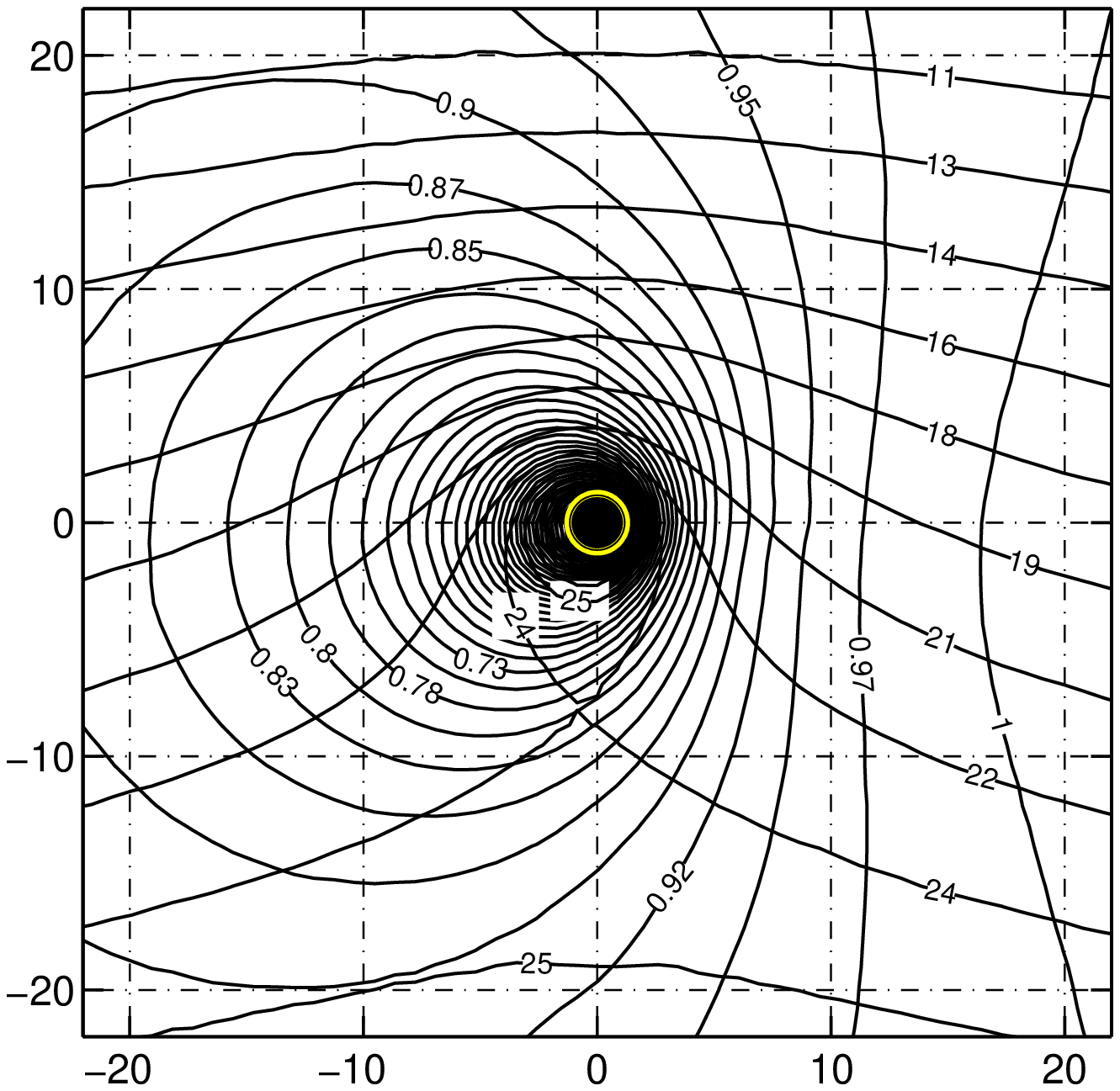}
\end{center}
\caption{Levels
of light-travel time $t(r,\phi)={\rm{}const}$ (approximately horizontal
direction of the contour lines) are plotted together with levels of the
redshift function $g(r,\phi)={\rm{}const}$ (roundish shape of the
latter). The contours are constructed in the equatorial plane of Kerr
black hole for two cases: a non-rotating hole ($a=0$, left panel) and for
a maximally rotating hole ($a=1$, right panel). Observer is located
towards top of the figure (at the inclination of
$\theta_{\rm{}o}=20^{\rm{}o}$). The argument of this paper assumes that
the spiral pattern crosses $t={\rm{}const}$ contours and the signal
varies thanks to their deformation and thanks to lensing near the
horizon. Geometrical units are used for time $t$ (conversion to physical
units as in the previous figure caption); redshift function $g$ is
dimensionless and it attains values around unity ($g>1$ corresponds to
the blue-shift, i.e.\ observed energy of photons higher than the
rest-frame energy). Three circles are plotted around the center: the
horizon radius ($r=r_{\rm{}h}$, black),  the circular photon orbit
($r=r_{\rm{}ph}$, yellow), and the marginally stable orbit
($r=r_{\rm{}ms}$, red). The circles coincide with each other in the
extremely rotating case: $r_{{\rm{}h}|a=1}=r_{\rm{}g}$.}
\label{fig2}
\end{figure}

Let us assume that a perturbation of local emissivity structure develops
on length-scales of $\simeq10$--$20r_{\rm{}g}$ extending along a
logarithmic spiral pattern $r\,\equiv\,r(\phi)$ (Karas et al.\ 2001).
The spirals become active either by their intrinsic synchrotron emission
and Compton up-scattering or the illumination from a primary source.
Although our model is a phenomenological one, such kind of spirals are
expected to arise by several mechanisms in accretion discs: spiral waves
represent large-scale structures (size comparable with the radius) that
can be induced by non-axisymmetric instability mechanisms.\footnote{Most
of the attention towards gaseous spiral waves has been originally
motivated by studies of cataclysmic variables. It has been recognized
that the variation of the density profile and of the ionization
structure of accretion flows, predicted by numerical and semi-analytical
methods, is followed by temperature modulation and, therefore, a change
in the gas (thermal) emissivity within the spirals. A similar effect is
expected for the X-ray irradiated accretion flows in AGN. Sanbuichi,
Fukue \& Kojima (1994) first considered the spirals extending close to a
Schwarzschild black hole and they showed examples of relativistically
distorted spectra where the effects of general relativity play a role.}
Also, a pattern resembling a single-armed spiral is produced from an
extended spot after its decay due to shearing (Karas, Vokrouhlick\'y \&
Polnarev 1992) or it may be produced by debris in the wake of a tidally
captured and disrupted satellite (e.g.\ Gomboc \& \v{C}ade\v{z} 2005).

One should emphasize that the physical conditions leading to the
spiral-shaped source of X-rays must be very different from those
envisaged by the orbiting spot model, although, on the phenomenological
level the two models do not appear to be that much different (they can
be treated by similar numerical schemes). The spot model has been
built on the standard disc, which is illuminated by coronal flares;
otherwise it remains almost intact. Spots are merely the reflection
features that come into existence only by flares and they cease as soon
as the irradiation is diminished. On the other hand, the existence of
extended spirals probably requires that the base flow is
non-axisymmetric and, hence, profoundly different from a standard disc
-- self-gravity, external forcing by another body, or MHD processes
must be invoked to create the spirals and calculate their form. They
light up themselves by the synchrotron mechanism.

\subsection{Time-delay calculations and the signal enhancement}
We calculated the light-travel time from the equatorial disc around a
Kerr black hole to a distant observer. Apart from the central mass $M_{\bullet}$,
the situation is characterized by parameters $a$ (dimension-less black
hole spin) and  $\theta_{\rm{}o}$ (inclination angle).\footnote{We
employ standard notation for the Kerr spacetime in Boyer-Lindquist
coordinates and geometrized units ($c=G=1$; e.g.\ Misner, Thorne \&
Wheeler 1973). All lengths and times are made dimensionless by
expressing them in units of the typical mass of the central black hole. 
Radius is supposed to be greater than
the marginally stable orbit, i.e.\ $r_{\rm{}ms}=3r_{\rm{}g}$ for a
non-rotating black hole $(a=0),$ and $r_{\rm{}ms}=1r_{\rm{}g}$ for a
maximally rotating black hole $(a=1)$.} The complexities of primary
X-ray reprocessing can be hidden by parameterizing the emissivity in
the form of a logarithmic spiral wave. The emission region extent and
shape are then defined by the spiral-wave pitch angle and the emissivity
contrast -- two variables that can be fitted to actual data. 

Adopting the phenomenological approach does not merely hide the unknown
physics. It also allows us to distinguish the principal difference of
the two models, i.e.\ their geometry, while the ``physical'' models in
reality rely on a number of free input parameters that have to be set.

We first estimate the light-crossing time across the spiral-wave extent.
It comes out of the order of $t_{\rm{c}}\approx10M_6$\,sec. On the other
hand, the orbital, thermal, sound-crossing, and viscous time-scales are
typically longer than $t_{\rm{}c}$. Radiation arrives at the observer
from different regions of the source, so that individual light rays
experience variable time lags. Time intervals get longer very near to
the hole because of gravitational delays predicted by general
relativity, including the frame-dragging effect near a rotating black
hole, which we also take into account.

The geometrical time lag (along different rays) can be characterized by
the maximum value ${\Delta}t(r)$, which also indicates whether the
Euclidean formula gives a correct value of the light-travel time with an
acceptable precision. Figure~\ref{fig1} shows ${\Delta}t$ for a source
located near $r=r_{\rm{}ms}$. Solid curves represent the delay values in
Kerr spacetime, while the dotted lines show the approximation in flat
space. Relativistic corrections are increasingly important for
$r\lesssim 5r_{\rm{}g}$, where ${\Delta}t(r)$ increases sharply. On the
other hand, the difference between the exact value of ${\Delta}t$ and
its Euclidean approximation is less than $10\,\%$ for a source location
$\gtrsim5r_{\rm{}g}$ (see Karas et al.\ 2001).

Figure~\ref{fig2} shows contours of relative time delay between a ray
coming from a given radius in the equatorial plane
($\theta=90^{\rm{}o}$), and an (arbitrarily chosen) reference ray. In
this figure, time delay was calculated in Kerr metric. Clearly, the
contours are progressively deformed and even split as the emission
radius approaches the black-hole horizon. Reference values quoted with
the contours of this figure can be transformed to physical time units
(measured by a distant observer) by the relation
$\bar{t}\,{\rm{}[sec]}\approx10M_6\,t$. Furthermore, contours of
constant redshift $g(r,\phi)={\rm{}const}$ are over-plotted in
Fig.~\ref{fig2}. Radiation flux is enhanced (or diminished) by factor
$g^4$ as it originates from the regions approaching (receding) the
observer.

Fig.~\ref{fig2} once again suggests the main grounds for the
enhancement of the observed signal. The enhancement occurs when 
photons emitted at different points of the rotating source
reach the observer similar time. This is possible near the black hole
($r\lesssim6r_{\rm{}g}$), where $t(r,\phi)={\rm{}const}$ contours are
bent significantly. The actual shape of the spiral supporting the 
signal enhancement depends also on the pattern rotation, i.e., not
solely on the spacetime geometry. Needless to say, the effect combines
with the lensing and Doppler amplification as the source crosses the
lensing caustics in $g>1$ region. 

Obviously the effect grows with spin of the black hole and attains
maximum at $a=1$, a theoretical upper limit for Kerr black hole. The
difference from the canonical $a=0.998$ case is rather minimal, except
for a small shift of $r_{\rm{}ms}(a)$ radius. That shift can prove to be
important for the disc emission though, provided that the inner edge of
the disc is attached to $r=r_{\rm{}ms}$.

\section{Discussion}
As mentioned above, the interplay of lensing and the Doppler boosting
was discussed by many authors within the orbiting spot model, whereas
the influence of time-delays has not been emphasized to such detail. The
effect is noticeable when watching the well-known animations of an
orbiting spot at a large view-angle inclination (e.g.\ Fig.~3 in Eckart
et al.\ 2007): the signal is sharply enhanced at the moment when the
large spot moves behind the black hole. In the case of a spiral-shaped
emission region the effect is expected to be even more pronounced thanks
to the elongated size of the source. 

The amplification is not symmetrical between leading and trailing
spirals of otherwise the same geometry and the intrinsic emissivity. Put
in a different way, the timing properties of the flare lightcurves can
in principle constrain the ratio of $v_r/v_\phi$ of the spiral pattern
producing them. In particular, for $v_r=0$, $v_\phi=v_{\rm{}K}$ the
model is effectively reduced to the orbiting spot model, whereas for
$l={\rm{}const}$ (constant angular momentum of the gas), $v_r<0$,
$v_\phi<v_{\rm{}K}$ the case goes over to the falling spot model. Very
exciting is now the possibility of having an extended source which can
be incorporated within the spiral model. On the other hand the effects
of lensing and the delay amplification should not be so important
in the case of a low-angular momentum inflow ($v_\phi\ll v_{\rm{}K}0$),
which has been also widely applied in the context of Sgr~A* (Proga \&
Begelman 2003; Moscibrodska et al.\ 2007, and references cited therein).

Further, it has been recognized that relativistic effects can strongly
influence the observed signal and enable us to measure physical
parameters of Sgr~A* black hole. The simultaneous near-infrared and
X-ray flares as well as the steady microwave emission from Sgr~A* may be
important probes of the gas dynamics and space-time metric of the black
hole. The enhancement of the signal discussed in the present paper
should be seen in all wavelengths as long as the approximation of
geometrical optics is satisfied.

We have argued that the emitting region is likely to be twisted into a
shape more complex than a simple spot. The spiral pattern is a
physically sound possibility for the flaring region, in which the effect
of relativistic modulation is more pronounced. The enhancement of the
main peak of the lightcurve takes place roughly on time-scale of the
spiral pattern crossing the equal-time curves. In other words the
duration of the event can be significantly shorter that the pattern
rotation period (it depends on the spiral shape and its rotation law).
On the other hand, the pattern orbital speed is still relevant for the
estimation of the flare periodicity over the entire cycle.

Given a specific mechanism to generate the spiral waves, certain freedom
remains in the model parameters, so the actual form of the spiral
profile can vary. Because for an ideal geometrical alignment of the
spiral a rather sharp enhancement of the observed signal is foreseen
(i.e., stronger than the spot model would predict), we can expect
occasional strong flares with amplitudes exceeding the more frequent and
currently known flares from Sgr~A*. 

\section{Conclusions}
Albeit physically substantiated, the model of an extended emission
region suffers from a practical disadvantage in comparison with the spot
model. The spiral model is more complex and the number of parameters
describing the source is greater. Therefore, the fitting procedure will
need better quality of future data. The spiral model assumes mechanisms
beyond the standard disc scheme play a major role and form these
non-axisymmetric structures. This may or may not be true. After all, the
two scenarios -- spots versus spirals -- can be relevant for different
categories of objects and different regimes of accretion. To this
uncertainty refers the question mark in the title of the paper. The
advent of simultaneous X-ray and IR detections of Sgr~A* flares and the
improving temporal and polarimetric resolution offer a promising
potential to remove ambiguities that still hamper the association
between physical models and real data.

\ack
The Czech authors thank the ESA PECS program (ref.~98040) and the
Center for Theoretical Astrophysics (ref.~LC06014) for the continued support.


\end{document}